\documentclass{ws-procs9x6}

\begin{document}

\title{Neutron Transversity at Jefferson Lab}

\author{J. P. Chen}

\address{Jefferson Lab, Newport News, VA 23606, USA \\
E-mail: jpchen@jlab.org}

\author{X. Jiang}

\address{Rutgers University, Piscataway, NJ 08855, USA}

\author{J.-C. Peng, L. Zhu}

\address{University of Illinois, Urbana, IL 61801, USA}

\author{for the Jefferson Lab Hall A Collaboration}

\maketitle

\abstracts{
Nucleon transversity and single transverse spin asymmetries have been the
recent focus of large efforts by both theorists and experimentalists.
On-going and planned experiments from HERMES, COMPASS and RHIC are mostly 
on the proton or the deuteron. Presented here is a planned measurement of the 
neutron
transversity and single target spin asymmetries at Jefferson Lab in Hall A
using a transversely polarized $^3$He target.  
Also presented are the results and plans of other neutron transverse spin 
experiments at Jefferson Lab.
Finally, the factorization for semi-inclusive DIS studies at 
Jefferson Lab is discussed.}

\section{Introduction}
\par
After forty years of extensive experimental and theoretical efforts, the 
unpolarized Parton Distribution Functions (PDFs) have been extracted 
from DIS, Drell-Yan and other processes with excellent precision over a 
large range of $x$. The comparison of the structure functions in a large 
range of $Q^2$ with QCD evolution equations have provided one of the best tests
of QCD. Since the ``proton spin crisis'' in the 1980s, 
very active spin-physics 
program have been carried out at CERN, SLAC and HERA, and,
recently, at JLab and RHIC. Longitudinally polarized parton distribution functions have been extracted by a number of groups in recent years, although the precision is not as good as that of the unpolarized PDFs.
The other equally important parton distribution functions, the transversity
distributions, have only been explored recently. 

\subsection{Transversity}
\par
The transversity distributions, $\delta q(x,Q^2)$,
are fundamental leading-twist (twist-2) quark distributions, same as the unpolarized
and polarized parton distributions, $q(x,Q^2)$ and $\Delta q(x,Q^2)$. 
In quark-parton models,
they describe the net transverse
polarization of quarks in a transversely polarized nucleon\cite{BDR}.
There are several special features for the transversity distributions,
making them uniquely interesting: 

\begin{itemize}
    \item
The difference between the transversity and the longitudinal distributions
is purely due to relativistic effects. In the absence of relativistic
effects (as in the non-relativistic quark model, where boosts and rotations
commute), the transversity distributions are identical
to the longitudinally polarized distributions.

    \item
          The quark transversity distributions do not mix with gluonic
effects\cite{BST} and therefore follow a much simpler evolution and have
          a valence-like behavior.

    \item The positivity of helicity amplitudes leads to the 
Soffer's inequality for the transversity\cite{soffer}:
    $\vert h_1^q \vert \le \frac {1}{2} (f_1^q + g_{1}^q)$.

    \item The lowest moment of $h_1^q$ measures a simple local
    operator analogous to the axial charge, known as the ``tensor charge'',
which can be calculated from lattice QCD.
\end{itemize}

Due to the chiral-odd nature of the transversity distribution, it can not
be measured in inclusive DIS experiments. In order 
to measure $\delta q(x,Q^2)$,
an additional chiral-odd object is required, such as double-spin asymmetries in
Drell-Yan processes, single target spin azimuthal asymmetries in
Semi-Inclusive DIS reactions, double-spin asymmetries in $\Lambda$ production
from e-p and p-p reactions and single-spin asymmetries in double pion 
production from e-p scattering.
The first results, from measurements performed by the HERMES\cite{hermes} and COMPASS\cite{compass} collaborations with SIDIS offered the first glimpse of possible effects caused by the transversity distributions.
 
\subsection{Semi-Inclusive Deep-Inelastic-Scattering}
\par
Semi-inclusive DIS is a powerful tool for probing various
parton distributions and fragmentation functions. For producing a spin-zero 
meson, the SIDIS differential cross section at leading
order contains 8 structure functions\cite{mulders96,mulders98}.
The single transverse target spin asymmetry term contains the Collins term\cite{collins},
which contains the product of the transversity and the Collins fragmentation 
function and is proportional to $\sin(\phi_h^\ell+\phi_S^\ell)$, the Sivers term\cite{sivers}, which contains the product of the 
Sivers structure function with the regular fragmentation function and is 
proportional to $\sin(\phi_h^\ell-\phi_S^\ell)$, and a third term proportional 
to $\sin(3\phi_h^\ell-\phi_S^\ell)$.
Azimuthal angles are defined relative to the lepton plane, 
e.g.\ $\phi_h^\ell = \phi_h
-\phi^\ell$ is the angle between the hadron plane 
and the lepton plane and  $\phi_S^\ell = \phi_S -\phi^\ell$ is the angle between
the spin plane and the lepton plane. 
SIDIS can be used to extract all 
eight of the twist-2 structure functions by choosing different beam helicity
and target polarization configurations,  
and detecting the
azimuthal angular dependences ($\phi_h^l$ and $\phi_s^l$).

\subsection{Experimental Status}
\par
The HERMES collaboration reported two years ago the observation of
single-spin azimuthal asymmetries
for charged and neutral hadron electroproduction\cite{hermes03}. Using an
unpolarized positron beam on longitudinally polarized hydrogen and deuterium
targets, the cross section was found to have a sin$\phi_h^l$ dependence.
Although a longitudinally polarized target was used in the HERMES experiment, 
there is a small ($\approx 0.15$) nonzero value of polarization transverse to
the virtual photon direction. 
The observed azimuthal asymmetries arise from Collins, Sivers and 
higher-twist contributions from the longitudinal target polarization. 
For a longitudinally
polarized target ($\phi_s^l = 0$) the Collins and the Sivers 
mechanisms can not be distinguished.

If the azimuthal asymmetry observed by HERMES is indeed partially caused by the
nucleon transversity, a much larger asymmetry is expected for a transversely
polarized target. 
The HERMES~\cite{hermes} and COMPASS~\cite{compass} collaborations have 
collected polarized SIDIS data using transversely polarized hydrogen and
LiD targets, respectively. A simultaneous fit to sin$(\phi_h^l + \phi_s^l)$
and sin$(\phi_h^l - \phi_s^l)$ dependences was applied to the HERMES data
to extract the ``Collins moments" and the ``Sivers moments",
respectively. Nonzero Collins moments were clearly observed
in the HERMES data with positive azimuthal asymmetry for $\pi^+$ and 
negative azimuthal asymmetry for $\pi^-$. The unexpectedly large magnitude
for the azimuthal asymmetry of $\pi^-$ relative to $\pi^+$ seems to suggest
that the disfavored Collins function is of comparable magnitude, but
opposite sign, to the favored Collins function. Furthermore, the observation
of the sin$(\phi_h^l - \phi_s^l)$ moments shows that the Sivers function is
nonzero and indeed contributes to the azimuthal asymmetry in SIDIS.
Preliminary results\cite{compass} from the COMPASS 2002 data show that the 
sin($\phi_h^l + \phi_s^l)$ azimuthal asymmetry is consistent with zero 
for a transversely polarized LiD target. A factor of 4 increase in
statistics is expected for the COMPASS 2002-2004 data.

To be able to extract the transversity from the Collins moments, independent
measurements of the Collins fragmentation functions are needed. First 
results of an extraction
of the Collins fragmentation functions from $e^+ + e^-$ measurements by the 
Belle collaboration at KEK is available now\cite{belle}.
  
\section{A planned measurement of neutron transversity at JLab}
\par
The Thomas Jefferson National Accelerator Facility (Jefferson Lab, or JLab)
produces a continuous-wave electron beam
of energy up to 6 GeV. An energy upgrade to 12 GeV is planned in the next 
few years. The electron beam with a current up to 180 $\mu$A is 
polarized up to $85\%$ by illumating a 
strained GaAs cathode with polarized laser light. The electron beam goes
into three experimental halls (Halls A, B and C) where the electron scattering
off various nuclear targets takes place. The experiments reported here studied 
semi-inclusive (or inclusive) 
electron scattering where the scattered electron and one scattered hadron
(or only teh scattered electron) are detected.
The neutron experiments presented here are from
Hall A\cite{NIMA} where a polarized $^3$He target, with in-beam polarization of about $40\%$,
provides an effective polarized neutron target. The polarized luminosity reached is $10^{36}$ s$^{-1}$cm$^{-2}$.
There are two High Resolution Spectrometers (HRS) 
with momentum up to 4 GeV/c. 
The HRS detector package consisted of vertical
drift chambers (for momentum analysis 
and vertex reconstruction), scintillation counters (data acquisition 
trigger) and particle identification detectors: gas \v{C}erenkov counters and lead-glass calorimeters for electron detection or Aerogel and heavy gas 
\v{C}erenkov detectors and RICH detectors for hadron particle identification. 
In addition to the HRS's, the BigBite spectrometer with a large acceptance is 
used for
detecting electrons for the semi-inclusive experiments where large acceptance
is needed. The BigBite detector package consists
of drift chambers, scintillators and a lead-glass calorimeter. 

A recently approved JLab proposal~\cite{e03004} plans to measure the 
single-spin asymmetry of the ${\vec n}(e,e^\prime \pi^-)X$ reaction on 
a transversely polarized $^3$He target.
The goal of this experiment is to provide the first measurement 
of the neutron transversity, complementary to the ongoing HERMES and COMPASS 
measurements on the proton and deuteron.
This experiment focuses on the valence quark region, $x=0.19 - 0.34$,
at $Q^2=1.77 - 2.73$ GeV$^2$.  
Data from this experiment, when combined with data from 
HERMES and COMPASS, will provide powerful constraints on the transversity
distributions of both $u$-quarks and $d$-quarks in the valence region.

\begin{figure}
\center {
\resizebox{0.7\textwidth}{!}{
  \includegraphics{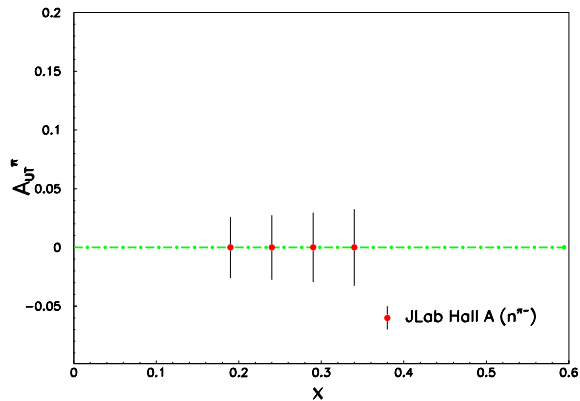}
}
\caption{
Expected statistical precision of this experiment 
%in comparison with HERMES run-II projected precision on a
%proton target~{\protect\cite{korotkov01}}
.}
}
\label{fig:1}  
\end{figure}

The experiment will use a 6 GeV electron beam with the Hall 
A left-side high resolution 
spectrometer (HRS$_L$) situated at 16$^\circ$ as the hadron arm,
and the BigBite spectrometer located at $30^{\circ}$ beam-right
as the electron arm.
A set of vertical coils will be added to the polarized $^3$He target 
to provide tunable polarization directions in all three dimensions. By
rotating the target polarization direction in the transverse plane, 
the coverage in $\phi_s^l$
is increased, hence facilitating the separation of the Collins
and the Sivers effects.
Figure~\ref{fig:1} shows the expected statistical precision of
this experiment with 24 days of beamtime for ${\vec n}(e,e^\prime \pi^-)X$ 
single spin asymmetry.
%together with the projected precision\cite{korotkov01} of HERMES run--II of
%the ${\vec p}(e,e^\prime \pi^+)X$ reaction.
Due to the good particle identification in the HRS, $K^-$ data will be 
collected at the same time, providing a set of precision data to study
the transverse spin asymmetries for semi-inclusive $K^-$ production.

The same measurement can be performed with the HRS spectrometer in positive polarity
to detect $\pi^+$ and $K^+$. A new proposal is being developed for this 
measurement.

\section{Inclusive neutron transverse spin experiments at JLab} 
\par

In addition to the transverse spin SIDIS experiments to study transversity, 
inclusive transverse spin experiments can be used to study higher-twist effects.
There are several completed inclusive (double) transverse spin experiments\cite{e97103,e99117,e94010} which precisely
measured the second spin structure function $g_2$ and its moment $d_2$.
$g_2$, unlike $g_1$ and $F_1$, can not be
interpreted in a simple quark-parton model. To understand $g_2$ properly, 
it is best to start with the operator product expansion (OPE) method.
In the OPE, neglecting quark masses, $g_2$ can be cleanly separated into a
twist-2 and a higher-twist term:
  \begin{eqnarray}g_2(x,Q^2) = g_2^{WW}(x,Q^2) +g_2^{H.T.}(x,Q^2)~.
  \end{eqnarray}
The leading-twist term, $g_2^{WW}$, can be determined from 
$g_1$\cite{WW}
and the higher-twist term arises from the quark-gluon correlations.
Therefore $g_2$ provides a clean way to study higher-twist effects.
In addition, at high $Q^2$, the $x^2$-weighted moment, $d_2$, 
is a twist-3 matrix element and is related to the color 
polarizabilities\cite{d2}.
Predictions for $d_2$ exist from various models and lattice QCD.

A precision measurement of g$_2^n$ from JLab E97-103~\cite{e97103} covered 
the Q$^2$ range from 0.58 to 1.36 GeV$^2$ at x $\approx 0.2$. 
Results for $g_2^n$ are given on the left panel of 
Fig. 2. The light-shaded area in the plot 
gives the leading-twist contribution, obtained by fitting world data\cite{BB} and
evolving to the $Q^2$ values of this experiment. The systematic errors are 
shown as the dark-shaded area near the horizontal axis.   

\noindent
\begin{figure}[!ht]
\parbox[t]{0.29\textwidth}{\centering\includegraphics[bb=660 22 1112 425, angle=-90,width=0.29\textwidth]{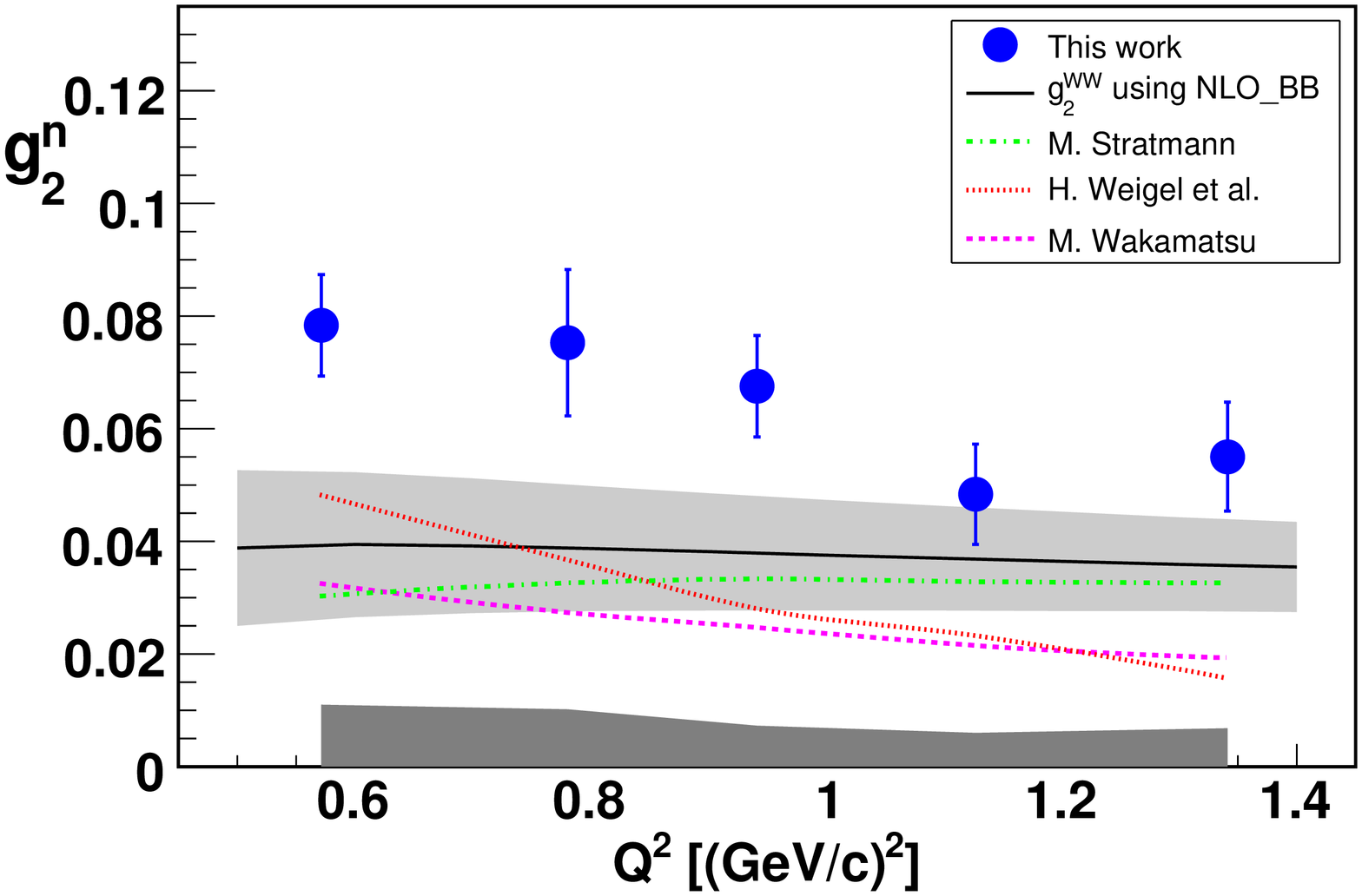}}
\parbox[t]{0.35\textwidth}{\centering\includegraphics[bb=-190 70 152 455, angle=0,width=0.35\textwidth]{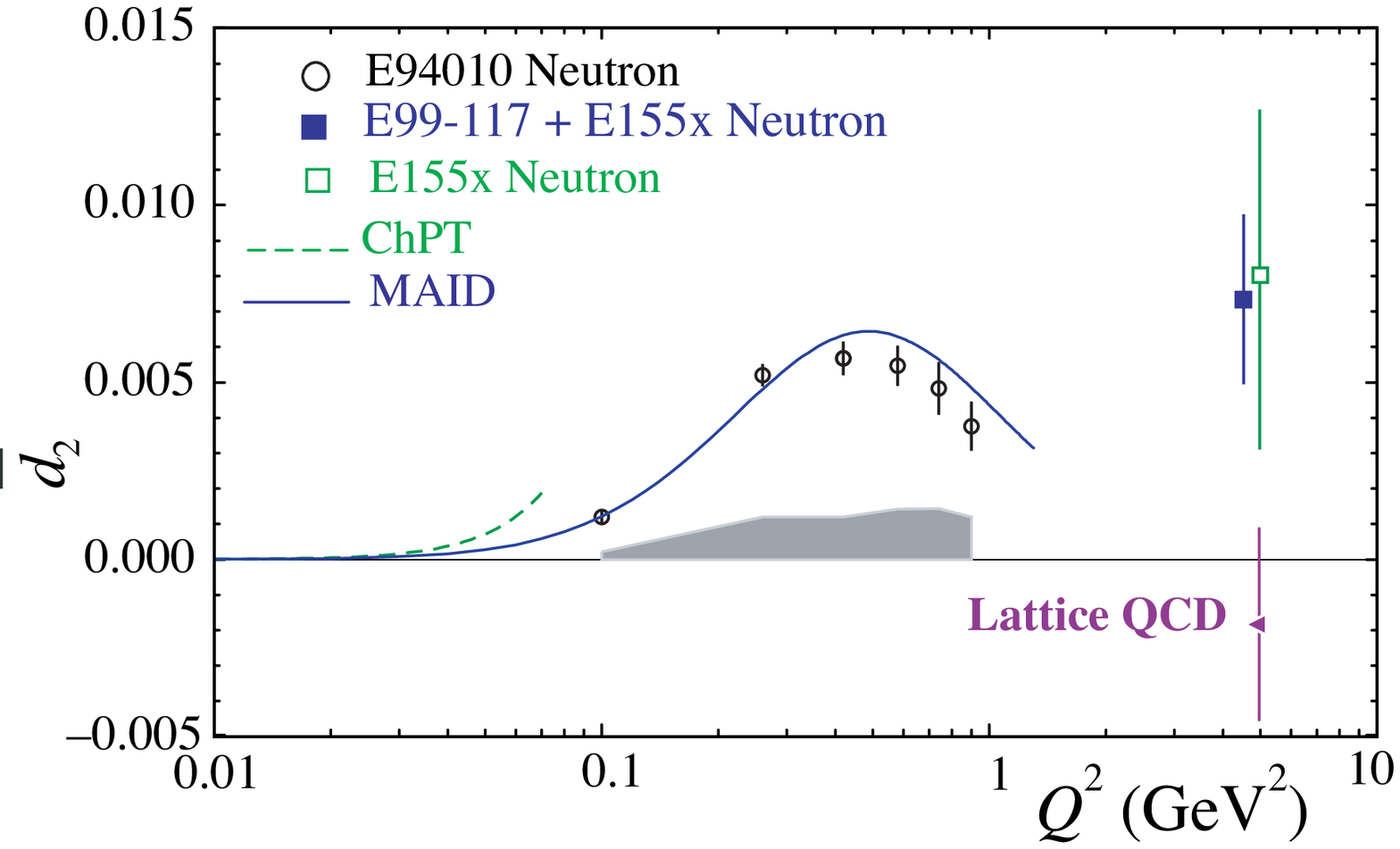}}
\vspace {-4cm}
\caption{Results for $g_2^n$ (left) and $d_2^n$ (right) from JLab Hall A.}
\end{figure}

The precision reached is more than an order
of magnitude improvement over that of the best world data\cite{E155x}.  The difference 
of $g_2$ from the leading twist part ($g_2^{WW}$) is due to 
higher twist effects. 
The measured $g_2^n$ values 
are consistently higher than $g_2^{WW}$.
For the first time, there is a clear indication that higher-twist effects 
become significantly positive at $Q^2$ below 1 GeV$^2$, 
while the bag model\cite{str} and Chiral Soliton model\cite{wei,wak} 
predictions of higher-twist effects are negative or close to zero. 

The second moment of the spin structure function, $d_2$,
can be extracted from $g_1$ and $g_2$ measurements.
Due to $x^2$ weighting, the contributions are dominated by the high-$x$ region and the problem of low-$x$ extrapolation is avoided. 
The Hall A experiment E99-117\cite{e99117} provided data on $A_2^n$ at high-$x$. 
 Combining these results with the world data\cite{E155x}, the second moment $d_2^n$ was extracted at an average $Q^2$ of 5
GeV$^2$.
This result is compared to the previously published result\cite{E155x} and
a calculation by Lattice QCD\cite{LQCD}.  
While a negative or near-zero value was 
predicted by Lattice QCD and most models, the new result for $d_2^n$ 
is positive. Also shown in Fig. 2 are the low $Q^2$ (0.1-1 GeV$^2$) results of the inelastic part of $d_2^n$ from another Hall A experiment E94-010\cite{e94010,chen05}, which were
compared with a Chiral Perturbation Theory calculation\cite{chpt} and a model prediction\cite{maid}.

\section{Experimental Tests of Factorization for SIDIS at JLab}
\par

Due to the limitation of maximum beam energy (6 GeV now and 12 GeV after the
planned energy upgrade), how well factorization works for SIDIS at JLab
is an important issue. Due to the high luminosity, it is possible to 
select kinematical settings keeping  $Q^2$ reasonably large by going to 
large scattering angles. With an optimal choice of kinematics, the typical 
SIDIS measurements at JLab will be at $Q^2$ around 2 GeV$^2$ with $W^2$ of 4-10
GeV$^2$ and $W'^2$ of around 4 GeV$^2$ for an $x$ range of 0.1-0.4 and 
a $z$ range
of 0.4-0.6. At what precision level factorization will work can only 
be answered by experimental tests. First test results are becoming available
from $p(e,e'\pi^{+,-})$ experiments at JLab Hall C\cite{Cfactor} and Hall B\cite{Bfactor}. These results
show that at the $10\%$ level, the data are consistent with the factorization
assumption.   

A JLab Hall A proposal~\cite{pr04114} has been conditionally approved to 
measure unpolarized 
$(e, e^\prime \pi^\pm)$ and $(e, e^\prime K^\pm)$ reactions. The new data
will provide further precision tests of factorization. In addition, 
the pion 
SIDIS measurement aims to determine $\bar d - \bar u$ with 
much better statistical accuracy than the existing HERMES data~\cite{hermes98},
which will provide a complementary measurement of the sea asymmetry 
to the Drell-Yan process\cite{garvey01}.

\section*{Acknowledgments}
This work is supported by the U.S. Department of Energy (DOE). The Southeastern University Research Associaition operates the Thomas Jefferson National Accelerator Facility for DOE under contract DE-AC05-84ER40150, Modification No. 175.

\end{document}